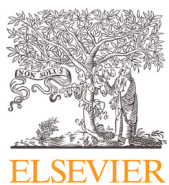
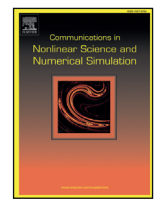

Research paper

# Basic principles drive self-organization of brain-like connectivity structure

Carlos Calvo Tapia [a,b], Valeri A. Makarov [a,c,∗], Cees van Leeuwen [b,d]

[a] *Instituto de Matemática Interdisciplinar, Universidad Complutense de Madrid, Madrid, Spain*
[b] *Laboratory for Perceptual Dynamics, Faculty of Psychology and Educational Sciences, KU Leuven, Leuven, Belgium*
[c] *N.I. Lobachevsky State University of Nizhny Novgorod, Nizhny Novgorod, Russia*
[d] *Center for Cognitive Science, TU Kaiserslautern, Kaiserslautern, Germany*

a b s t r a c t

The brain can be considered as a system that dynamically optimizes the structure of anatomical connections based on the efficiency requirements of functional connectivity. To illustrate the power of this principle in organizing the complexity of brain architecture, we portray the functional connectivity as diffusion on the current network structure. The diffusion drives adaptive rewiring, resulting in changes to the network to enhance its efficiency. This dynamic evolution of the network structure generates, and thus explains, modular small-worlds with rich club effects, features commonly observed in neural anatomy. Taking wiring length and propagating waves into account leads to the morphogenesis of more specific neural structures that are stalwarts of the detailed brain functional anatomy, such as parallelism, divergence, convergence, super-rings, and super-chains. By showing how such structures emerge, largely independently of their specific biological realization, we offer a new conjecture on how natural and artificial brain-like structures can be physically implemented.

## 1. Introduction

From early development throughout adulthood, brain anatomy is continuously altered by the mechanisms of structural plasticity [1]. We generally consider these mechanisms to have a role in optimizing the functional connectivity, i.e., the efficiency of communication across the anatomical networks [2,3].

For a given anatomical network, functional connectivity can be predicted by stochastic simulations in terms of graph diffusion [4]. Graph diffusion then drives the structural plasticity of the network according to the principle of adaptive rewiring [5], establishing shortcut connections between units that communicate intensively, while pruning underused connections [6]. Adaptive rewiring models capture the basic principle of how network structures evolve dynamically, in accordance with their spontaneous communication patterns [2,7].

Adaptive rewiring eventually leads to a small-world graph structure [7] with modular [2,8] and rich club organization [9]. The modular small-world structure [10] and rich-club effect [11] are pervasive properties of biological neural network

---

structure. In adaptive rewiring models, these properties emerge independently of whether functional connectivity is realized, e.g., through oscillatory synchronization [2,6–9] or timed spiking [12]. The generic emergence of these properties may contribute to their widespread character in the brain. Beyond biological systems, adaptive rewiring may be considered as a mechanism for optimizing the structure of artificial systems, such as deep learning networks [13] or neuroanimats [14].

Adaptive rewiring can furthermore take into account wiring cost [15], which can bias adaptive rewiring among network units localized in 2- or 3- or n-dimensional spaces. Minimizing wiring distance optimizes adjacency in neural network topography [16]. Note that the factoring of wiring costs in adaptive rewiring models does not remove long-range connections [15]. The system retains a sparse but stable proportion of long-range connections. These typically develop into hub connections, whereas the short-range connections largely become intermodular. Such a differentiation emerges gradually, analogously to how hubs develop in brains [17]. Thus, small-world architectures emerge, in which the modules are located in adjacent regions, with sparse long-range intermodular connections. The spatial mapping of the modules reflects the macroscopic characteristics of brain anatomy [18].

However, adaptive rewiring so far has failed to account for the emergence and variety of more detailed functional neuronal architectures. For instance, retinal ganglion cells typically have a concentric structure allowing local convergence of the neuronal activity [19]. Elsewhere, pools of neurons form super-chains [20] allowing volleys of activity to propagate across the cortex as synfire chains or polychronous sets [21]. We will show that such properties can indeed emerge from adaptive rewiring.

To this end, we assume that neuronal activity is associated with a potential field [22]. Early in development, propagating fields play a role in shaping neural architecture. Brain activity around gestation is characterized by massive bursts of action potentials that bring into synchrony neighboring cells and spread in a wave-like manner (for a review, see [23]). In response to this activity, axons grow, synaptic boutons are formed and relayed. During adulthood, more subtle factors such as traveling waves characterize brain activity, both within [24] and across [25–27] cortical regions. Both in the cortex and hippocampus, these waves play a role in establishing structural plasticity, in particular during sleep (see, e.g., [28] and references therein). These studies mostly investigated the experience-dependent plasticity of selected synapses for structural consolidation. The current model, however, is concerned with the nonspecific effects of intrinsic wave activity. The principle of spike-timing-dependent plasticity prescribes that long-term potentiation occurs when presynaptic inputs lead to postsynaptic spikes [29]. In analogy, we assume that connections are established in the direction of wave propagation.

To take this function into account, it is sufficient to bias the rewiring according to the direction of propagating waves of spontaneous network activity. We consider the effect of some of the most frequently observed wave patterns. Radial, planar, and rotating waves are observed in thalamocortical and intracortical networks. Which form predominates may depend on factors like connectivity, the balance between excitation and inhibition, and transmission delays between thalamus and cortex [30]. Here we are interested in the effect of these waves, rather than their origin. We will compare the effect of concentric and linearly propagating waves superimposed on the network dynamics. The resulting network dynamics is seen to preserve the evolution of modular small-world structure and wiring length minimization while shaping the morphology of concentric ganglions and super-chains.

## 2. Results

### 2.1. Network rewiring principles

We consider graphs describing networks of $n$ neural units linked by $m$ connections (see Methods). The units may represent single neurons or, rather, populations of mixed inhibitory and excitatory neurons of different morphologies, metabolisms, and response types. For simplicity, we assume that the nodes in a graph are located on a unit disk (Fig. 1A, left, for $n = 7$).

The networks are adaptively rewired as follows. Initial arbitrary connectivity evolves over time as connections are pruned while new ones are created according to three rewiring principles (see Methods). Principle P1: *distance* (Fig. 1A.1) represents wiring cost optimization. At each cycle, the longest connection is removed and replaced by a connection between the closest unconnected pair of units. Principle P2: *field* (Fig. 1A.2) prescribes rewiring according to a propagating wave field. It removes the connection closest to orthogonal to the propagation direction of the wave field and replaces it with one, closest to parallel with the propagation direction between a previously unconnected pair of units. Together, Principles 1 and 2 change the network based on spatial criteria. Principle P3: *graph diffusion* (Fig. 1A.3) reflects rewiring in adaptation to network function [6]. This adaptation involves removing an underused connection, while establishing a connection between a hitherto unconnected pair of units that has the most intensive traffic between them, as represented by the graph diffusion.

The principles P1–P3, applied in isolation, have relatively trivial effects on the network structure. Fig. 1B illustrates them over an initially randomly wired network (leftmost). P1 (distance) removes long-distance couplings, substituting them by shorter ones until the network topology stabilizes (Fig. 1B.1). P2 (field) applied to the same random network with lateral waves oriented along the *y*-axis imposes a prevalence of vertically oriented links (Fig. 1B.2). The impact of P3 (diffusion) on the topological reorganization is little, at the first glance (Fig. 1B.3). However, using functional representation, which minimizes the number of crossing links, we observe a significant functional modification of the network (Fig. 1B.3, inset). It has been fallen apart into several densely interconnected substructures.

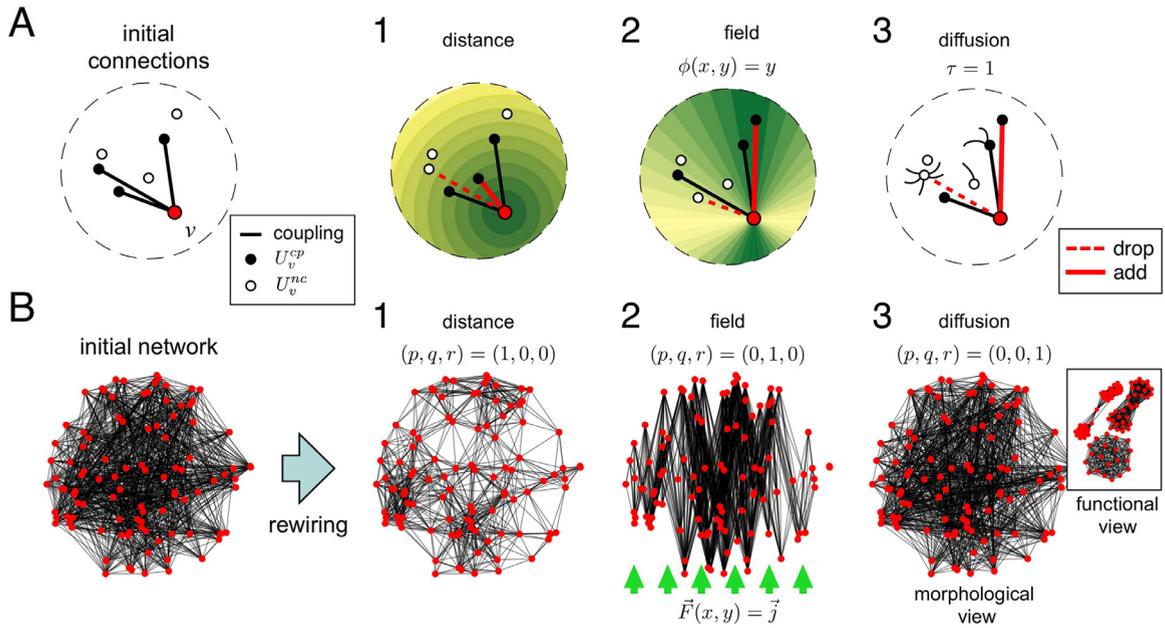

**FIG. 1.** Basic principles driving structural evolution of small neural networks. (A) Left: A network consisting of 7 neurons randomly distributed on a unit disk (only the couplings with neuron $v$, red circle, are shown). Black circles mark neurons coupled with $v$ (set $U_v^{cp}$) and open circles correspond to neurons not coupled with $v$ (set $U_v^{nc}$). Right: Three principles of network rewiring. P1: by distance (1), P2: by external field (2), and P3: by graph diffusion (3) [diffusion parameter $\tau = 10$; background color varies from green to yellow according to the criterion value related to neuron $v$; dashed and solid red links indicate connections dropped and added, respectively]. (B) Left: Initially random network ($n = 100$, $m = 912$). Right: Examples of the network structures obtained by applying the rewiring principles taken separately after 18, 240 cycles. All networks are shown in morphological view, i.e., the topographical distances are faithfully represented. The inset on the right provides a functional view of the network after applying P3 [the nodes are rearranged in space to minimize the number of the crossing connections]. (For interpretation of the references to color in this figure legend, the reader is referred to the web version of this article.)

### 2.2. Probabilistic rewiring due to P1–P3: emergence of super-chains

The effectivity of the three principles resides in their combination. Whereas each of them is deterministic, the choice of the principle applied at each rewiring step is probabilistic. The triplet ($p$, $q$, $r$), such that $p + q + r = 1$, describes the weights (probabilities) at which each principle P1, P2, or P3 is chosen. The above described effects (Fig. 1B) were obtained assuming either $p = 1$, or $q = 1$, or $r = 1$. Let us now study the mutual impact of the rewiring principles P1–P3.

Since P3 (diffusion) is the only functional criterion, in further simulations, we always use it, i.e., $r > 0$. Taken P3 separately ($r = 1$), we observed an artifact, not found in biological neural networks: the network loses connectivity and falls apart (Fig. 1B.3, inset). As soon as P3 is applied in combination with any of the other two rules (i.e., $r < 1$), the connectivity between the nodes persists in time and modular small-world structures emerge. Earlier, such a result was achieved by combining P3 with a random rewiring at each step [6]. Here, however, we show that the random rewiring is not necessary.

Fig. 2A shows an initially random network with $n = 100$ nodes and $m = 912$ couplings (topologically identical to that shown in Fig. 1B, left). We now apply to this network 3648 rewiring cycles by combining principles P1–P3. The network resulting from the combination of P1 with P3 (Fig. 2B) inherited the properties found earlier [6,15]. On the one hand, P1 successfully reduces the wiring length (see also Fig. 1B.1) and on the other hand, the network gains a modular structure due to P3 (Fig. 2B, inset). As we will show below, this network has a small-world structure.

The application of P2 and P3 again drives the network to a small-world structure (Fig. 2C). Here, however, the connections are topographically oriented along the wave, as expected. One then could make a conclusion that P1 and P2 have similar functional effects. Surprisingly, if all three principles are active, a novel nonlinear effect is obtained (Fig. 2D). While anatomically the network structure does not differ significantly from what one could expect from the structures obtained earlier (Figs. 2B–D, main panels), in the functional space a superstructure emerges (Fig. 2D, inset). We thus found a novel phenomenon: formation of a super-chain consisting of sparsely coupled modules of clustered neurons, widely observed in brain networks. Thus, besides the individual effects, combination of all rewiring principles brings an unexpected emerging property.

### 2.3. Effects of wave on the emerging network structures

Let us now test how external waves affect the evolution of neural networks. Fig. 3 compares the evolution in case of a lateral wave (similar to Fig. 2D, but with different parameters) and a radial wave. Again, we observe that the lateral wave organizes modules in a functional superstructure, in this case in a super-ring (Fig. 3B, inset). Only few shortcut links break the ring geometry. The chain shown in Fig. 2D and the ring in Fig. 3B may serve as the structural basis for a synfire chain [20] or a polychronous set [21], on which volleys of synchronized activity propagate.

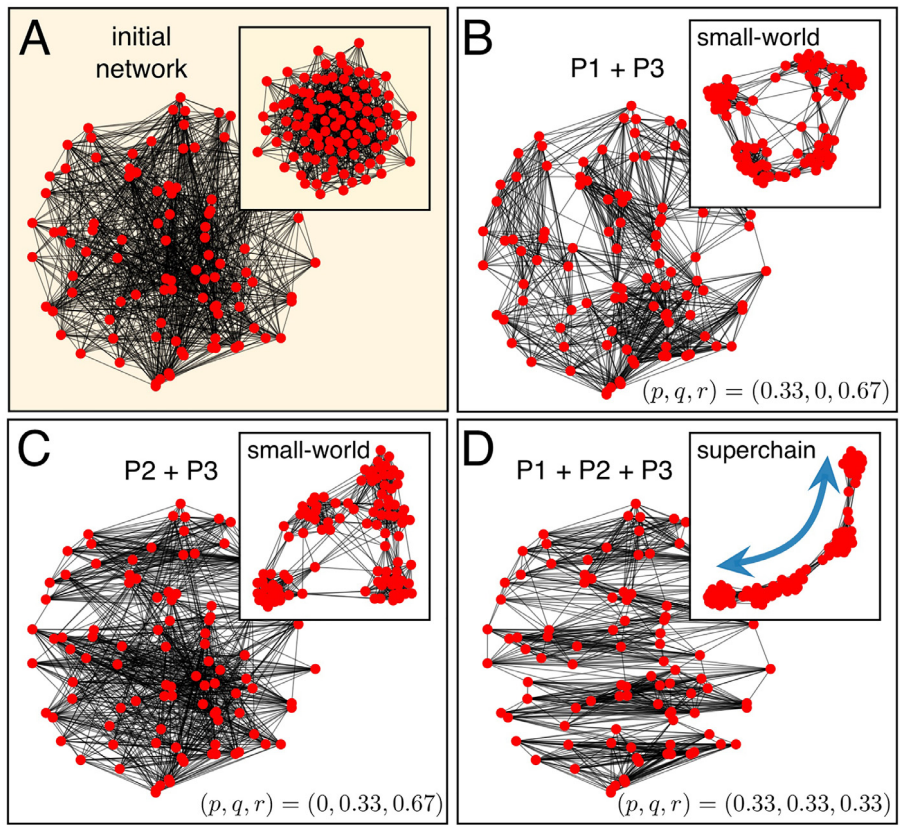

**FIG. 2.** Combination of P1–P3 drives neural network to different anatomical (main panels) and functional (insets) states. (A) Initial random neural network ($n = 100$, $m = 912$). (B) Effect of P1 (distance) and P3 (diffusion, $\tau = 1$). The wiring cost is reduced and the network gains a modular small-world structure (inset). (C) Effect of P2 (external lateral wave propagating along $x$-axis) and P3. A modular small-world structure is similar to that shown in (B). (D) All three criteria working together. A functional super-chain emerges (inset) [the lateral wave as in (C)].

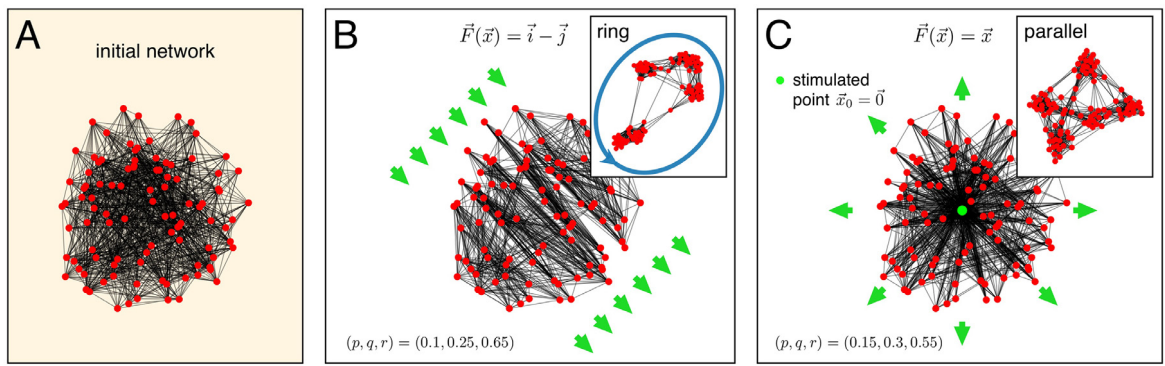

**FIG. 3.** Effect of external waves on formation of superstructures ($n = 100$, $m = 912$, $\tau = 1$). (A) Initial randomly connected neural network. (B) Stabilized network structure after 3648 rewiring cycles formed by a lateral wave. (C) Same as in (B) but with a radially propagating wave.

In case of a radial wave (Fig. 3C), a ganglion is formed with two clusters in the center of the topography. The superstructure consists of a set of parallel clusters connected to the ones in the center, which may serve as inputs and outputs (Fig. 3C, inset).

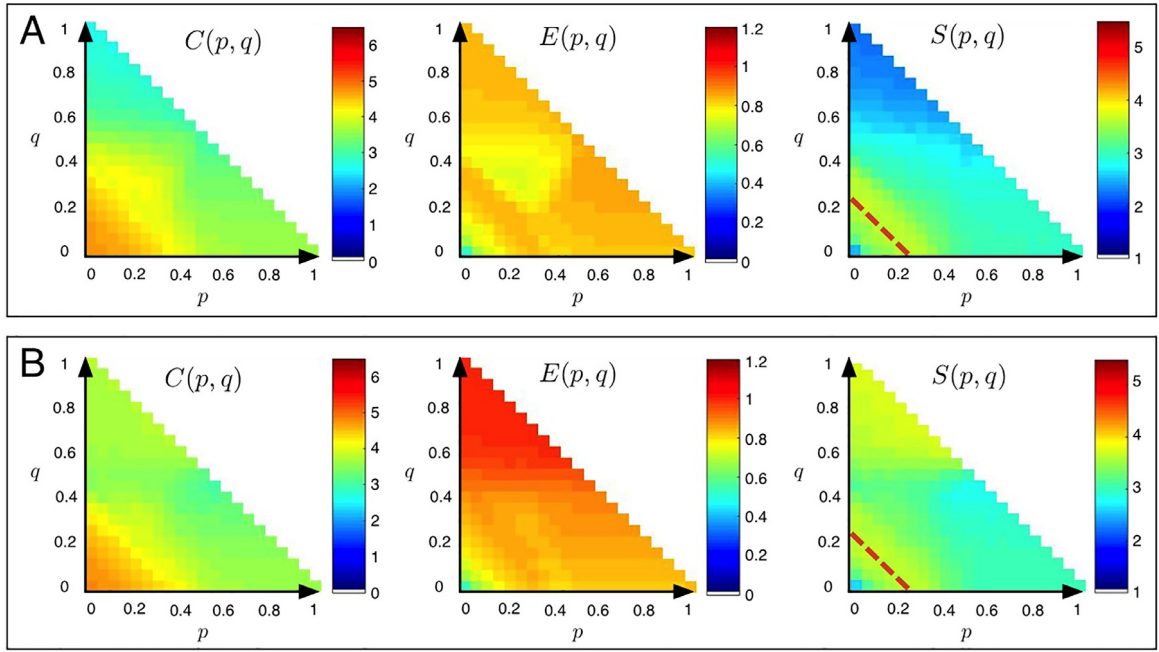

**FIG. 4.** Mean structural indexes for different combinations of principles P1–P3 [Left: the normalized cluster coefficient *C*; Middle: the normalized global efficiency *E*; Right: the small-world index *S*. $\tau = 1$]. (A) Network rewiring under a lateral field. (B) Network rewiring under a radial field.

*2.4. Small-world structure*

In the examples shown above, we stipulated that a small-world structure emerges under network evolutions driven by rewiring principles P1–P3. Let us now substantiate this claim by evaluating small-worldness indices [6] for all combinations of the probabilities *p, q*, and *r* describing the weights of P1–P3, respectively.

We use the normalized small-world index:

$$S = \frac{C}{C_r} \frac{E}{E_r},$$

where *C* is the global clustering coefficient, *E* is the global efficiency of the network [3], while $C_r$ and $E_r$ are the corresponding measures for an equivalent Erdös-Renyi (ER) random network [31]. For a random network, we have $C \approx C_r$, $E \approx E_r$ and hence $S \approx 1$. Then, the higher *S*, the greater the degree of small-worldness.

Fig. 4A shows the normalized average indexes *C, E*, and *S* as a function of the rewiring probabilities *p* and *q* for both laterally (as in Fig. 3B) and radially propagating waves (as in Fig. 3C). In both cases, the highest levels of small-worldness are achieved for the same proportions of *p* and *q* (dashed red lines in Fig. 4), mainly because the largest gains in clustering are made in these regions, with only small losses in efficiency. At the maxima we have

$$q + p = 0.23,$$

which then corresponds to $r = 0.77$ (dependent on $\tau$). This shows that the clustering is mainly driven by the graph diffusion. In addition, there is a region that benefits specifically from a large contribution of radial waves, which affects, in particular, the global efficiency (Fig. 4B, middle panel). This result may be ascribed to a centralizing effect induced by those fields, as observed in Fig. 3C.

## 3. Discussion

Adaptive rewiring generates networks that are universally small-worlds. Thus, it captures the common principle of how these structures emerge in the brain through structural plasticity. By taking rewiring costs into account, the resulting networks are brain-like in their macroscopic topography. As we have shown, these characteristics are preserved when networks are rewired according to the propagating direction of a wave field. The wave field, however, adds a crucial component to the emergence of the functional anatomy characteristic of the brain. Networks exposed to waves propagating laterally develop structures consisting of pools of densely interconnected neurons that form super-chains through intermodular connections. The presence of such chain structures was a necessary condition for the formation of synfire chains in realistically scaled neuronal models [32,33]. Networks exposed to radially propagating waves show the emergence of ganglia, supporting parallelism with a divergent input and a convergent output.

The present results illustrate that key aspects of the neuronal network organization can be obtained with a few very basic self-organization principles. This makes the existence of such functional organization in biological networks a robust phenomenon of their dynamics, rather than a product of gene expression. To study the emergence of functional anatomy in the brain, self-organization should be considered as pre-eminent, whereas gene expression may set the parameters under which these networks evolve, such that the prominence of certain structures may differ from one region to another.

Certain brain diseases, such as schizophrenia, are today considered dysconnectivity diseases [34]. Under certain parameter regimes, our model can produce suboptimal connectivity [35]. Thus, it may have a heuristic function to help explain how dysconnectivity can arise.

Optimization of the network structure is a common problem in artificial systems, such as deep learning networks. The adaptive rewiring principles P1–P3 apply straightforwardly to such systems. Rewiring can be applied during downtime, in analogy to what happens in the human brain during sleep [28,36].

## 4. Methods

To describe the network structure, we introduce an undirected graph $G = (V, E)$ with $n$ vertices or neurons, $V = \{1, 2, ..., n\}$, and $m$ edges or couplings, $E = \{(i, j) \in V^2 : i \text{ linked with } j\}$. Then, for each neuron $v \in V$ we define two subsets (Fig. 1A, left): (1) neurons coupled to $v$ and (2) neurons not coupled to $v$:

$$U_v^{cp} = \{j \in V \setminus \{v\} : (j, v) \in E\} \\ U_v^{nc} = \{j \in V \setminus \{v\} : (j, v) \notin E\} \quad (1)$$

The network structure changes through rewiring of couplings. At each iteration, we select a neuron $v \in V$ and then rewire its couplings. The rewiring occurs in accordance with one of the three organizing principles.

**P1:** *Close neurons wire together*. We drop the longest edge within the set $U_v^{cp}$ and add the shortest link from the set $U_v^{nc}$ (Fig. 1A.1).

Principle P1 corresponds to the experimentally validated observation that the spatially close neurons have higher probability of being coupled. This principle reduces the wiring cost [37].

To model the effect of waves of neuronal activity widely observed in the brain [38], we introduce a vector field orthogonal to the wave front as a gradient of a potential field:

$$\vec{F}(\vec{x}) = \nabla \varphi(\vec{x}). \quad (2)$$

For illustration purpose, we use two examples: 1) Lateral activation: $\varphi = y$ and then $\vec{F} = \vec{j}$, i.e., the field is directed along $y$-axis (Fig. 1A.2); 2) Radial activation: $\varphi = \frac{x^2+y^2}{2}$ and then $\vec{F} = x\vec{i} + y\vec{j}$, and we get a radially oriented field (Fig. 3C). Thus, we formulate the second principle.

**P2:** *Couplings tend to orient along waves*. We introduce an external vector field oriented orthogonally to the wave front. Then, we drop the coupling closest to orthogonal to the field from the set $U_v^{cp}$ and add one closest to parallel from the set $U_v^{nc}$ (Fig. 1A.2).

Principle P2 has also experimental and theoretical background. Stimulation of a neural network produces traveling waves of excitation that propagates outwards the stimulus location. Such waves shape couplings among neurons through STDP by potentiating those that are aligned with the wave direction [39].

To model the effect of functional diffusion, we introduce the third principle. Its application involves the creation of shortcuts in "overloaded" environments (with intensive "diffusion") while underused connections are annihilated.

**P3:** *Functional diffusion reroutes traffic*. This principle optimizes traffic of information among neurons by rerouting it through new paths. For example, couplings with heavily coupled neurons disappear while links with less loaded neurons appear (Fig. 1A.3).

The formal description of P1-3 is as follows. Given a neuron $v \in V$ we define the sets $U_v^{cp}$, $U_v^{nc}$ [Eq. (1)] and

$$j_- = \underset{i \in U_v^{cp}}{\mathrm{argmax}}\{m_{iv}\}, \quad j_+ = \underset{i \in U_v^{nc}}{\mathrm{argmin}}\{m_{iv}\}, \quad (3)$$

where $j_-$ and $j_+$ are the indexes of the neurons losing the link and gaining a link with neuron $v$, respectively. In Eq. (3) the sets $\{m_{iv}\}$ are elements of a matrix $M = (m_{ij})_{i,j} \in V$ that depends on the used principle. We now introduce the interneuron distance matrix:

$$D = (d_{ij})_{i,j \in V}, \quad d_{ij} = \|\vec{x}_i - \vec{x}_j\|. \quad (4)$$

Then, principle P1 is given by Eq. (3) with $M = D$.

To introduce principle P2, we set a vector field $\vec{F}(\vec{x})$ and introduce the effective wave matrix:

$$W = (w_{ij})_{i,j \in V}, \quad w_{ij} = 1 - |\cos \theta_{ij}|, \quad (5)$$

where $\theta_{ij}$ stands for the angle between the vector defining the link from neuron $i$ to $j$ ($\vec{x}_j - \vec{x}_i$) and the field $\vec{F}(\vec{x}_i)$ with convention $\theta_{ii} = 0$. The angle can be evaluated as usual by:

$$\cos\theta_{ij} = \frac{\langle \vec{x}_j - \vec{x}_i, \vec{F}(\vec{x}_i) \rangle}{d_{ij} \|\vec{F}(\vec{x}_i)\|}, \quad i \neq j. \tag{6}$$

Thus, matrix $W$ defines the collinearity of couplings and the field (direction of wave). Then, P2 is given by Eq. (3) with $M = W$.

To introduce the third principle, we need some definitions. Let $A = (a_{ij})_{i,j \in V}$ be the adjacency matrix of the graph $G$ ($a_{ij} = 1$ if $(i, j) \in E$, and $a_{ij} = 0$ otherwise) and $B = (b_{ij})_{i,j \in V}$ be the diagonal matrix of degrees of all vertices ($b_{ii} = deg(i)$, $b_{ij} = 0$, $i \neq j$). Then, we introduce the normalized Laplacian matrix:

$$\mathcal{L} = I_n - B^{-\frac{1}{2}} A B^{-\frac{1}{2}} \tag{7}$$

with the convention $b_{ii}^{-1} = 0$ for $b_{ii} = 0$. Finally, we define the one-parametric exponential heat kernel [6]:

$$H(\tau) = (h_{ij}(\tau))_{i,j \in V} = -e^{-\tau \mathcal{L}} = -\sum_{k=0}^{\infty} \frac{(-\tau)^k}{k!} \mathcal{L}^k \tag{8}$$

where we used negative sign for convenience. This kernel models propagation of heat or information flow in the network. Then, P3 is given by Eq. (3) with $M = H$.

We now postulate the following iterative process. The graph $G$ representing a neural network is progressively rewired, with probabilities $p$, $q$, and $r$ ($p + q + r = 1$) defining the weights of principles P1, P2, and P3, respectively, in accordance with the following algorithm.

- **S0**: Build an undirected graph $G = (V, E)$ with $n$ vertices and $m$ edges representing the initial state of a neural network and set all parameter values.
- **S1**: Choose a random vertex $v \in V$, such that $0 < deg(v) < n - 1$, and define the sets $U_v^{cp}$ and $U_v^{nc}$ [Eq. (1)].
- **S2**: Select randomly one of the three criteria of rewiring: P1 with probability $p$, P2 with probability $q$, or P3 with probability $r$. Update the edges $E$ by deleting $(v, j_-)$ and adding $(v, j_+)$ couplings, where $j_-, j_+$ are chosen by Eq. (3) with the corresponding matrix $M$.

Repeat steps S1 and S2 unit the network structure stabilizes.

We note that although the number of couplings $m$ can be arbitrary, it is convenient to set $m = \lfloor 2 \log(n)(n-1) \rfloor$. This allows keeping the connectivity index (mean vertex degree) in biophysically reasonable limits (e.g., for $n = 100$ we get $m = 912$ and the connectivity index is about 9). This ensures (with high probability) that the initial graph will be connected. The number of iteration of steps S1 and S2 are set to $4m$.

## Funding

This work was supported by the Russian Science Foundation (project 19-12-00394), by the Spanish Ministry of Science, Innovation and Universities (grant FIS2017-82900-P), and by the Flemish Organization for Science (Odysseus grant G.0003.12).

## Declaration of Competing Interest

The authors declare no conflict of interests.